\documentclass[pra,showpacs,twocolumn]{revtex4-1}
\usepackage{amsfonts}
\usepackage{amsmath}
\usepackage{amssymb,amscd}
\usepackage{psfrag}
\usepackage{graphicx}
\usepackage{braket}
\usepackage[dvips]{epsfig}
\usepackage{epsfig}
\usepackage{subfigure}
\usepackage{color}
\usepackage{amssymb}

\begin{document}

\title{The effect of Quantum Statistics on the sensitivity in an SU(1,1)
interferometer}
\author{Jie Zeng$^{1}$}
\thanks{These authors contributed equally to this work.}
\author{Yingxing Ding$^{1}$}
\thanks{These authors contributed equally to this work.}
\author{Mengyao Zhou$^{1}$}
\author{Gao-Feng Jiao$^{2}$}
\author{Keye Zhang$^{1,3}$}
\author{L. Q. Chen$^{1,3}$}
\email{lqchen@phy.ecnu.edu.cn}
\author{Weiping Zhang$^{3,4,5,6}$}
\author{Chun-Hua Yuan$^{1,3}$}
\email{chyuan@phy.ecnu.edu.cn}

\address{$^1$State Key Laboratory of Precision Spectroscopy, Quantum Institute for Light and Atoms,
	Department of Physics, East China Normal University, Shanghai 200062, China}
\address{$^2$School of Semiconductors and Physics, North University of
China, Taiyuan 030051, China}
\address{$^3$Shanghai Branch, Hefei National Laboratory, Shanghai 201315,
China}
\address{$^4$Department of Physics, Shanghai Jiao Tong University, and
Tsung-Dao Lee Institute, Shanghai 200240, China}
\address{$^5$Shanghai
Research Center for Quantum Sciences, Shanghai 201315, China}
\address{$^6$Collaborative Innovation Center of Extreme Optics, Shanxi
University, Taiyuan, Shanxi 030006, China}

\begin{abstract}
We theoretically study the effect of quantum statistics of the light field
on the quantum enhancement of parameter estimation based on cat state input
the SU(1,1) interferometer. The phase sensitivity is dependent on the
relative phase $\theta$ between two coherent states of Schr\"{o}dinger cat
states. The optimal sensitivity is achieved when the relative phase is $\pi$%
, i.e., odd coherent states input. For a coherent state input into one port, the phase sensitivity of the odd coherent state into the second input port is inferior to that of the squeezed vacuum state input.
However, in the presence of losses the Schr\"{o}dinger cat states are more resistant to loss than squeezed vacuum states. As the amplitude of Schr\"{o}dinger
cat states increases, the quantum enhancement of phase sensitivity
decreases, which shows that the quantum statistics of Schr\"{o}dinger cat
states tends towards Poisson statistics from sub-Poisson statistics or
super-Poisson statistics.
\end{abstract}

\date{\today }
\maketitle


\section{Introduction}

In quantum metrology, the Mach--Zehnder interferometer (MZI) and its variant
SU(1,1) interferometer \cite{Yurke86} had been used as a generic model to
realize precise measurement of phase. Using quantum measurement techniques
to beat the standard quantum limit (SQL) has received a lot of attention in
recent years \cite%
{Helstrom67,Holevo82,Caves81,Braunstein94,Braunstein96,Lee02,Giovannetti06,Zwierz10,Giovannetti04,Giovannetti11,Hudelist14}%
. By optimizing over all possible estimators, measurements, and probe states
the accuracy of parameter estimation can be improved. Forty years ago, Caves
\cite{Caves81} suggested to replace the vacuum fluctuations with the
squeezed-vacuum light in the MZI to reach a sub-shot-noise sensitivity. Soon
after, various quantum resources were used to improve the measurement
precision \cite{Xiao, Grangier, Dowling08, N00N}. Due to the even coherent
states also are superposition of even number states, they are closely
related to the squeezed vacuum states, but with different coefficients. The
even and odd coherent states is called also Schr\"{o}dinger cat states and
is written as $|cat\rangle =(|\alpha \rangle +e^{i\theta }|-\alpha \rangle )/%
\sqrt{N}$ with $N=2+2e^{-2\alpha ^{2}}cos\theta $ is the normalization
factor and $\theta $ is the relative phase, where $\theta =0$, $\pi $, and $%
\pi /2$ corresponds to the even coherent states, odd coherent states, and
Yurke-Stoler states \cite{Gerry05}. The even and odd coherent states have
been predicted in interferometric detection of gravitation waves for
reducing the optimal intensity of the input laser \cite{Ansari}. The photon
statistics of single-mode and multimode even and odd coherent states are
also studied \cite{Buzik92, Ansari94}. The use of non-classical states can
enhance the accuracy of parameter estimation, which comes from the quantum
statistics of non-classical states: super-Poissonian statistics or
sub-Poissonian statistics. Given a constraint on the amplitude of Schr\"{o}%
dinger cat state, analyzing the quantum enhancement of the parameter
estimation by changing the quantum statistics of the cat state is worth
studying to further explain the origin of the quantum enhancement of the
phase parameter estimation. On the other hand, the measurement precision of
interferometers will be reduced in the presence of environment noise \cite%
{Escher,Dem12,Dem09}. For interferometers, the photon losses is a typical
decoherence process to be taken into account. Analyzing and studying the
influence of photon loss on the sensitivity of the cat state, and comparing
the result with that of the squeezed vacuum state is very helpful for the
application of the cat state in quantum metrology.

Quantum parameter estimation theory establishes the ultimate lower bound to
the sensitivity with which a classical parameter (e.g., phase shift $\phi $)
that parameterizes the quantum state can be measured. The popular way to
obtain the lower bounds in quantum metrology, is to use the method of the
quantum Fisher information (QFI) \cite{Braunstein94,Braunstein96}. It
establishes the best precision that can be attained with a given quantum
probe. In recent years, to obtain the QFI many efforts have been made by
theory \cite{Toth, PezzBook, Demkowicz,Wang,Jarzyna, Monras, Pinel, Liu,
Gao,Jiang,Safranek,Sparaciari15,Sparaciari} and experiments \cite%
{Strobel,Hauke}. Since interferometers as a generic model to realize precise
estimation of phase shift, the corresponding QFI has been studied. For the
MZI, Jarzyna \emph{et al.} \cite{Jarzyna} studied the QFIs between phase
shifts in the single arm case and in the two-arm case and presented the
differences between them. Many measurement processes do not include phase
reference, and the conclusions of QFI-only calculations will be
overestimated, which cannot be realized at all via exact inputs and
measurements \cite{Jarzyna, Takeoka, Zhong, Zhang}. For phase shifts in the
two-arm case, the phase estimation can be considered as a two parameter
estimation problem. In general, the calculation of the quantum Fisher
information matrix (QFIM) is necessary \cite{Lang1, Lang2, Ataman, You19,
Zhong2}.

In this paper, we study the cat state injection SU(1,1) interferometer and
analyze the relationship between photon statistics and phase sensitivity. To
avoid overestimation from QFI-only calculation, the conclusive quantum Cra\'{%
m}er-Rao bound (QCRB) based on the QFIM approach is derived. The sensitivity
of the optimized cat state input and the squeezed vacuum state input are
compared and studied, as well as the loss-resistant behavior in the presence
of losses.

\section{Phase sensitivity}

\subsection{QFI and Mandel Q-parameter}

\begin{figure}[t]
\centerline{\includegraphics[width=0.45\textwidth,angle=0]{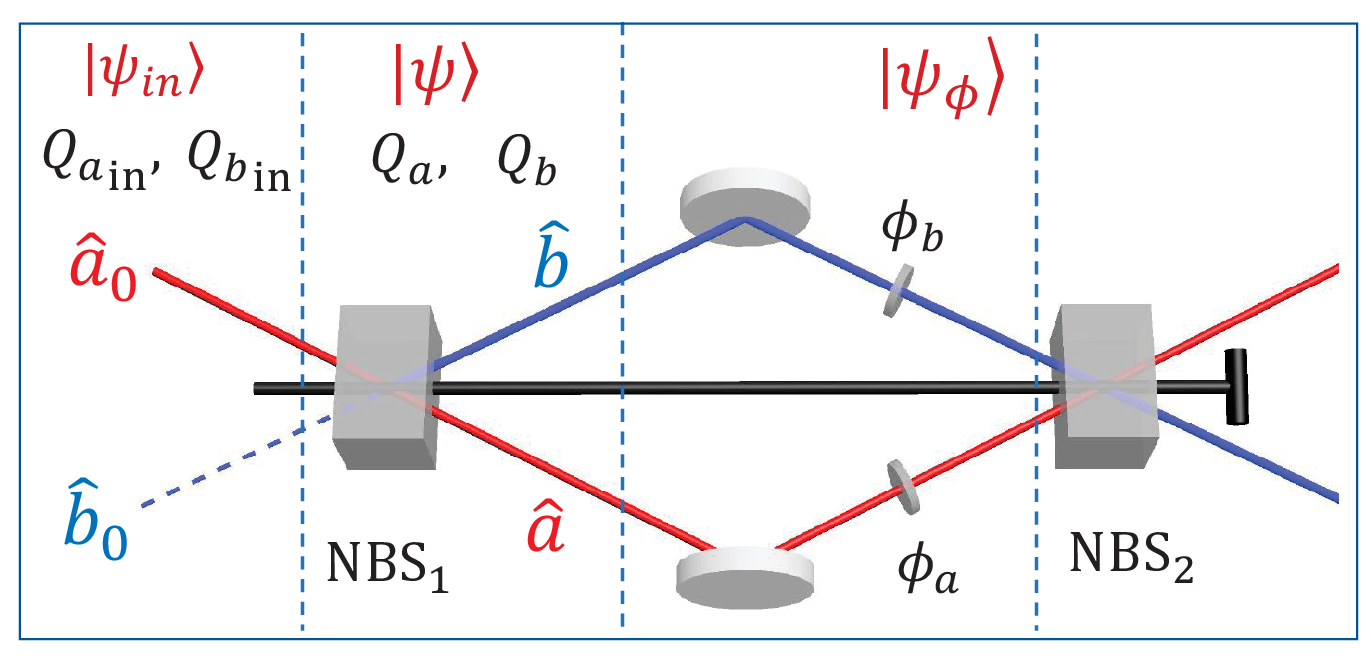}}
\caption{Schematic of an SU(1,1) interferometer. $|\protect\psi _{in}\rangle
$: initial state; $|\protect\psi \rangle $: probe preparation; $|\protect%
\psi _{\protect\phi }\rangle $: state after phase encoding. $\protect\phi %
_{a}$, $\protect\phi _{b}$: phase shift; $\mathrm{NBS}_{1,2}$: nonlinear
beam splitter; $Q_{a_{in},b_{in}}$, and $Q_{a,b}$ are the Mandel
Q-parameters of the input light fields and the light fields of the two arms
inside the interferometer, respectively.}
\label{fig1}
\end{figure}

An SU(1,1) interferometer is considered, as shown in Fig.~\ref{fig1}, where
the nonlinear beam splitters (NBSs) can be realized by the optical
parametric amplifiers (OPAs) or four-wave mixing (FWM) \cite{Jing11,Chen15}.
In the Heisenberg picture the transformation of the mode operators is
\begin{equation}
\hat{a}=u\hat{a}_{0}+v\hat{b}_{0}^{\dagger },\text{ \ }\hat{b}^{\dagger
}=u^{\ast }\hat{b}_{0}^{\dagger }+v^{\ast }\hat{a}_{0},
\end{equation}%
where $u=\cosh g$, and $v=e^{i\theta _{g}}\sinh g$. After the light field
passing the first NBS, as shown in Fig.~\ref{fig1}, the two beams sustain
phase shifts, i.e., mode $a$ undergoes a phase shift of $\phi _{a}$\ and
mode $b$ undergoes a phase shift of $\phi _{b}$. The transformation of phase
shift is described as
\begin{equation}
\hat{U}_{\phi }^{ab}=\exp [i\phi _{a}\hat{n}_{a}+i\phi _{b}\hat{n}_{b}]=\exp
(i\phi _{s}\hat{g}_{s})\exp (i\phi _{d}\hat{g}_{d}),
\end{equation}%
where $\hat{n}_{a}=\hat{a}^{\dagger }\hat{a}$, $\hat{n}_{b}=\hat{b}^{\dagger
}\hat{b}$, $\phi _{s,d}=\phi _{a}\pm \phi _{b}$ and $g_{s,d}=(\hat{n}_{a}\pm
\hat{n}_{b})/2$. In the SU(1,1) interferometer the quantity to estimate is
the phase sum $\phi _{s}$ \cite{Kok10}.

In general, the phase estimation as a two-parameter estimation problem, the
QFIM approach is necessary. For the estimation of $\phi _{s}$ and $\phi _{d}$
we can use the method of QFIM, which is given by a two-by-two matrix
\begin{equation}
\mathcal{F}=\left[
\begin{array}{cc}
F_{ss} & F_{sd} \\
F_{ds} & F_{dd}%
\end{array}%
\right] ,
\end{equation}%
where $F_{ij}=4(\langle \psi _{\phi }|\hat{g}_{i}\hat{g}_{j}|\psi _{\phi
}\rangle -\langle \psi _{\phi }|\hat{g}_{i}|\psi _{\phi }\rangle \langle
\psi _{\phi }|\hat{g}_{j}|\psi _{\phi }\rangle )$ ($i,j=s,d$). In the
SU(1,1) interferometer, we are usually interested in the estimation of only $%
\phi _{s}$ and its corresponding QFI is given by
\begin{equation}
\mathcal{F}=\frac{1}{\Delta ^{2}\phi _{s}}=F_{ss}-\frac{F_{sd}F_{ds}}{F_{dd}}%
.
\end{equation}%
After calculation, the $\mathcal{F}$ is\ worked as
\begin{equation}
\mathcal{F}=4\frac{\langle \Delta ^{2}\hat{n}_{a}\rangle \langle \Delta ^{2}%
\hat{n}_{b}\rangle -Cov[n_{a},n_{b}]^{2}}{\langle \Delta ^{2}\hat{n}%
_{a}\rangle +\langle \Delta ^{2}\hat{n}_{b}\rangle -2Cov[n_{a},n_{b}]},
\end{equation}%
where $\langle \Delta ^{2}\hat{n}_{i}\rangle =\langle \hat{n}_{i}^{2}\rangle
-\langle \hat{n}_{i}\rangle ^{2}$ ($i=a,b$), $Cov[n_{a},n_{b}]=\langle \hat{n%
}_{a}\hat{n}_{b}\rangle -\langle \hat{n}_{a}\rangle \langle \hat{n}%
_{b}\rangle $, $\langle \cdot \rangle $ denotes the average value $%
\left\langle \psi \right\vert \cdot |\psi \rangle $, and $|\psi \rangle $ is
the state of probe state $|\psi _{in}\rangle $ after being prepared by a
NBS. We use the Mandel $Q$-parameter to describe the photon statistics of
state $|\psi \rangle $, which is given by
\begin{equation}
Q_{i}=\frac{\left\langle \hat{n}_{i}^{2}\right\rangle -\langle \hat{n}%
_{i}\rangle ^{2}}{\langle \hat{n}_{i}\rangle }-1\text{ }(i=a,b),
\end{equation}%
where $\langle \Delta \hat{n}_{i}\rangle =\sqrt{\langle \Delta ^{2}\hat{n}%
_{i}\rangle }$. Using the Mandel $Q$-parameter, the $\mathcal{F}$ is\
rewritten as
\begin{equation}
\mathcal{F}=\frac{4(\left\langle \hat{n}_{a}\right\rangle \left\langle \hat{n%
}_{b}\right\rangle (Q_{a}+1)(Q_{b}+1)-Cov[\hat{n}_{a},\hat{n}_{b}]^{2})}{%
\left\langle \hat{n}_{b}\right\rangle (Q_{b}+1)+\left\langle \hat{n}%
_{a}\right\rangle (Q_{a}+1)-2Cov[\hat{n}_{a},\hat{n}_{b}]}.\text{ }\text{ }
\end{equation}%
Note that $Q_{a,b}$ here describes the Mandel $Q$-value of the light field in the two
arms of the interferometer.

The above inequality shows that the metrological advantage of nonclassical
light is primarily the photon statistics Mandel parameters $Q_{a}$ and $%
Q_{b} $ in the two arms of interferometer, and convariance $Cov[\hat{n}_{a},%
\hat{n}_{b}]$. Next, we study the effect of quantum statistics of input
fields on the sensitivity in an SU(1,1) interferometer.

\begin{figure}[t]
\centerline{\includegraphics[width=0.45\textwidth,angle=0]{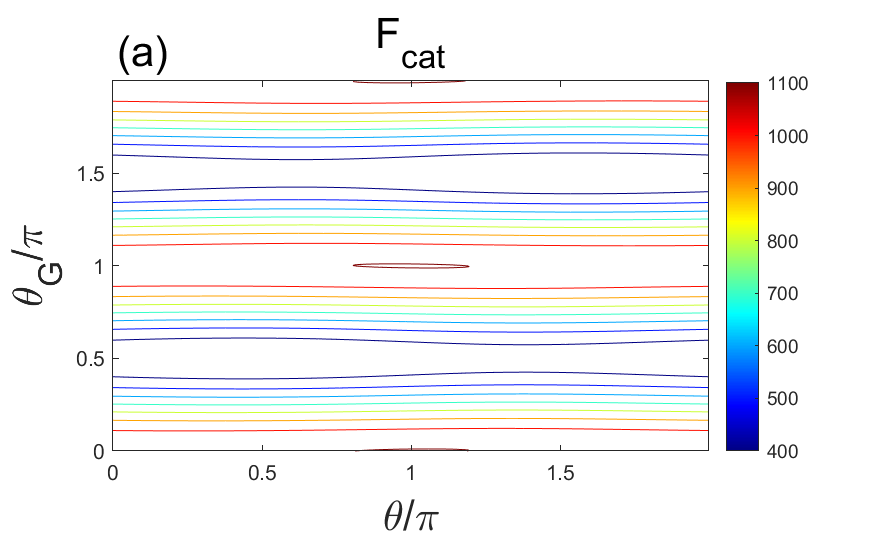}} %
\centerline{\includegraphics[width=0.45\textwidth,angle=0]{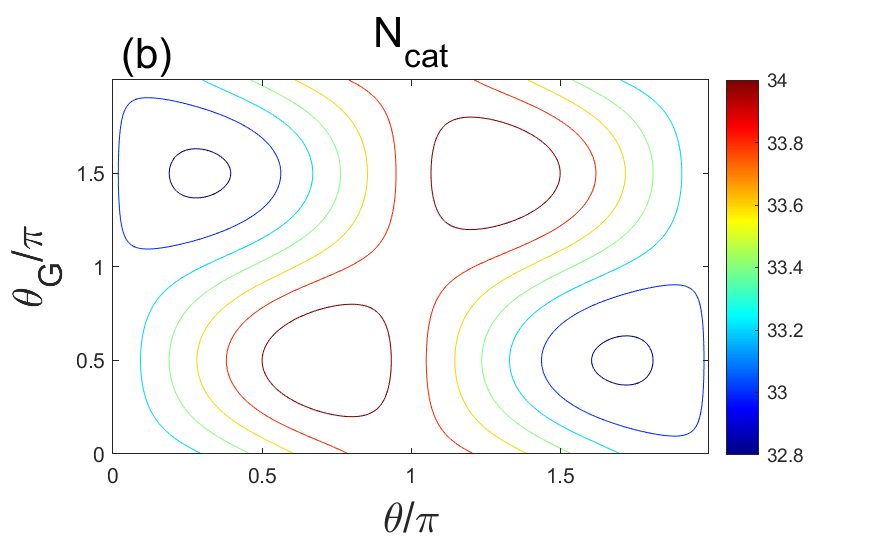}}
\caption{Schr\"{o}dinger cat state and coherent state input. (a) QFI $%
F_{cat} $ and (b) total number of photons $N_{cat}$ inside the
interferometer as a function of $\theta$ and $\theta_{G}$, where $g=1.2$, $\protect\alpha=1.2$, and $\protect\beta%
_{G}=1.8$.}
\label{fig2}
\end{figure}

\subsection{QFI of Schr\"{o}dinger cat state and coherent state}

We analysis for the input of Schr\"{o}dinger cat state $|cat\rangle
=(|\alpha \rangle +e^{i\theta }|-\alpha \rangle )/\sqrt{N}$ with $%
N=2+2e^{-2\alpha ^{2}}cos\theta $ in mode $a$ and coherent state $|\beta
\rangle $ ($\beta =\left\vert \beta \right\vert e^{i\theta _{\beta }}$) in
mode $b$. By using the QFIM method, we have
\begin{eqnarray}
\mathcal{F}_{\mathrm{cat}} &=&\sinh ^{2}(2g)[\bar{n}_{\alpha }+(1+2\bar{n}%
_{\alpha })\left\vert \beta \right\vert ^{2}+1]  \notag \\
+ &&4\cosh ^{2}(2g)\left\vert \beta \right\vert ^{2}+4\cosh (2g)\bar{n}%
_{q}+4(\bar{n}_{Q}-\bar{n}_{q}^{2})  \notag \\
- &&4\frac{[\cosh (2g)\left\vert \beta \right\vert ^{2}+(\alpha ^{2}+\bar{n}%
_{\alpha })\bar{n}_{q}]^{2}}{\alpha ^{4}+\bar{n}_{\alpha }-\bar{n}_{\alpha
}^{2}+\left\vert \beta \right\vert ^{2}},
\end{eqnarray}%
with
\begin{eqnarray}
\bar{n}_{q} &=&(e^{-2|\alpha |^{2}}\sin \theta /z_{s})\sinh (2g)\alpha
\left\vert \beta \right\vert \sin (\theta _{G}),  \notag \\
\bar{n}_{Q} &=&\frac{1}{2}\sinh ^{2}(2g)\alpha ^{2}\left\vert \beta
\right\vert ^{2}\cos (2\theta _{G}),
\end{eqnarray}%
where $\theta _{G}=\theta _{\beta }-\theta _{g}$, $\bar{n}_{\alpha }=\langle
cat|\hat{a}_{0}^{\dag }\hat{a}_{0}|cat\rangle =\alpha ^{2}(z_{d}/z_{s})$
with $z_{s,d}=1\pm e^{-2\alpha ^{2}}cos\theta $, $\bar{n}_{\alpha }$ is the
average photon number of cat state. Based on the QFI, the ultimate precision
bound of phase sensitivity is given by the QCRB
\begin{equation}
\Delta \phi _{\mathrm{QCRB}}=\frac{1}{\sqrt{m\mathcal{F}}},
\end{equation}%
where $m$ is the number of independent repeats of the experience.

When $\beta =0$, i. e., $|cat\rangle _{a}\otimes |0\rangle _{b}$ input, $%
\mathcal{F}_{\mathrm{cat}}$ is reduced to QFI of Schr\"{o}dinger cat state
and vacuum state
\begin{equation}
\mathcal{F}_{\mathrm{cat}}=\sinh ^{2}(2g)(1+\bar{n}_{\alpha }).  \label{QFIM}
\end{equation}%
The QFI of above Eq.~(\ref{QFIM}) is dependent on the average photon number
of cat state $\bar{n}_{\alpha }$. For a given amplitude $\alpha $ of Schr%
\"{o}dinger cat state, when the relative phases\ $\theta $ is changed, the
average number of photon $\bar{n}_{\alpha }$ is also changed. When $\theta
=0 $, $\pi $, and $\pi /2$, $\bar{n}_{\alpha }$ corresponds to $\alpha ^{2}%
\frac{1-e^{-2\alpha ^{2}}}{1+e^{-2\alpha ^{2}}}<\alpha ^{2}$, $\alpha ^{2}%
\frac{1+e^{-2\alpha ^{2}}}{1-e^{-2\alpha ^{2}}}>\alpha ^{2}$, and $\alpha
^{2}$. When $\theta =(2k+1)\pi $ ($k=0,1,2,\cdots $), the maximal $\bar{n}%
_{\alpha }$ is obtained for a given $\alpha $. When $\beta \neq0$, for a
given $\alpha $ and $|\beta |$, the QFI $\mathcal{F}_{\mathrm{cat}}$ as a
function of $\theta $ and $\theta _{G}$ is shown in Fig. 2(a). The maximum
value of QFI $\mathcal{F}_{\mathrm{cat}}$ occurs at $\theta =(2k+1)\pi $ and
$\theta _{G}=k\pi $ ($k=0,1,2,\cdots $). Thus, the optimal sensitivity is
achieved when odd coherent state input.

The quantum enhancement of phase sensitivity can be explained by the Mandel $%
Q$-parameter of input fields. The $Q$-parameter for the even coherent states
is $Q_{a_{in}}=4\alpha ^{2}e^{-\alpha ^{2}}/(1+e^{-4\alpha ^{2}})>0$
indicating super-Poissonian statistics, and for the odd coherent states is $%
Q_{a_{in}}=4\alpha ^{2}e^{-\alpha ^{2}}/(1+e^{-4\alpha ^{2}})<0$ indicating
sub-Poissonian statistics, and for Yurke-Stoler states is $Q_{a_{in}}=0$.
Since the odd coherent state input can obtain the optimal phase sensitivity,
that is, the input field with sub-Poisson statistics $(Q_{a_{in}}<0)$ can
obtain the best sensitivity for phase measurement, which is opposite to the $%
Q_{a}$ and $Q_{b}$ value of the light field of the two arms inside the
interferometer.

As $\alpha $ becomes large, $Q_{a_{in}}$ $\rightarrow 0$, which shows the
effect of Schr\"{o}dinger cat states is equivalent to coherent state. The
enhancement of phase sensitivity due to quantum statistics gradually
decreases to $0$ as $\alpha $ increases.

\begin{figure}[t]
\centerline{\includegraphics[width=0.45\textwidth,angle=0]{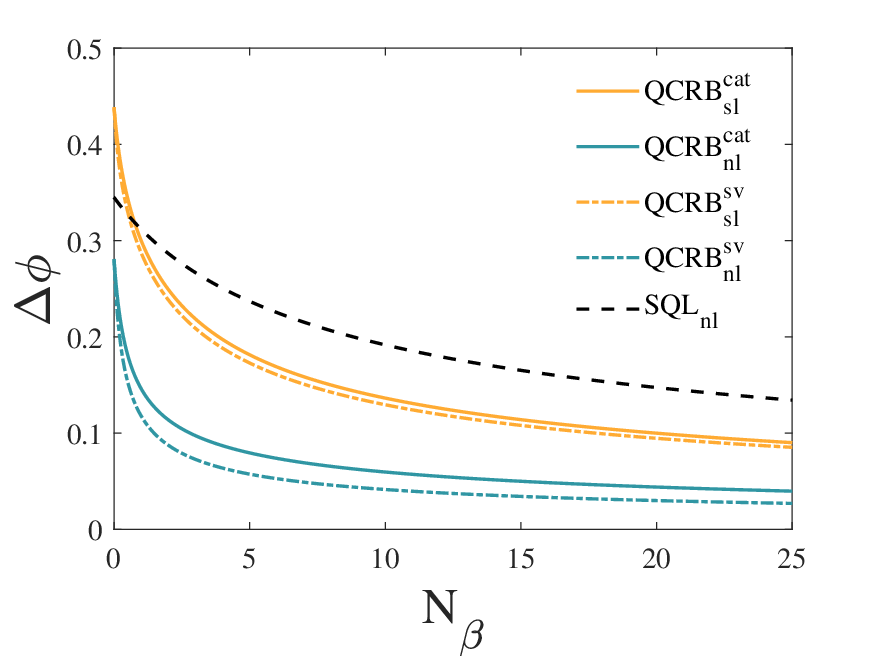}}
\caption{Phase sensitivity versus the photon number $N_{\protect\beta }$
with the two input states $\left\vert cat\right\rangle _{a}\otimes
\left\vert \protect\beta \right\rangle _{b}$ and $\left\vert \protect\xi %
\right\rangle _{a}\otimes \left\vert \protect\beta \right\rangle _{b}$ and
average photon number $\bar{n}_{\protect\alpha }=\bar{n}_{s}$, $\protect%
\alpha =2$, $g=1.2$, $\protect\eta _{a}=0.5$, $\protect\eta _{b}=1$, $%
\protect\theta =\protect\pi $ and $\protect\theta _{G}=\pi$. QCRB$_{p}^{k}$: $%
k\in $[cat, $\left\vert cat\right\rangle _{a}\otimes \left\vert \protect%
\beta \right\rangle _{b}$ case; sv, $\left\vert \protect\xi \right\rangle
_{a}\otimes \left\vert \protect\beta \right\rangle _{b}$ case], $p\in $[sl,
single arm loss case; nl, lossless case]. The black dotted line is the SQL
in the lossless case.}
\label{fig3}
\end{figure}

\section{Comparison between cat state and Squeezed state}

It is also worthwhile to compare the above result with that of squeezed
vacuum $|0,\xi \rangle $ $(\xi =re^{i\eta }$) in mode $a$ and coherent state
in mode $b$, where $r$ is the squeezing strength of the squeezed vacuum, and
its QFI is as follows~\cite{You19}:
\begin{eqnarray}
\mathcal{F}_{_{\mathrm{sv}}} &=&\sinh ^{2}(2g)[\left\vert \beta \right\vert
^{2}e^{2r}+\cosh ^{2}(r)]  \notag \\
&&+\cosh ^{2}(2g)\frac{8\left\vert \beta \right\vert ^{2}\sinh ^{2}(2r)}{%
4\left\vert \beta \right\vert ^{2}+2\sinh ^{2}(2r)}.
\end{eqnarray}%
When $\beta =0$, $\mathcal{F}_{\mathrm{sv}}$ is reduced to
\begin{equation}
\mathcal{F}_{\mathrm{sv}}=\sinh ^{2}(2g)[1+\bar{n}_{s}],  \label{Fsq}
\end{equation}%
where $\bar{n}_{s}=\sinh^{2}r$, $\bar{n}_{s}$ is the average photon number
of the squeezed vacuum state. Comparing Eq.~(\ref{QFIM}) with Eq.~(\ref{Fsq}%
), their QFIs are equal when $\bar{n}_{\alpha }=$ $\bar{n}_{s}$.

The total number of photons $\langle \hat{N}\rangle $ inside the SU(1,1)
interferometer is given by $\langle \hat{N}\rangle \equiv \langle \hat{a}%
^{\dag }\hat{a}\rangle +\langle \hat{b}^{\dag }\hat{b}\rangle $. For input
state $|0,\xi \rangle _{a}\otimes |\beta \rangle _{b}$ and $|cat\rangle
_{a}\otimes |\beta \rangle _{b}$, their total number of photons $N_{\mathrm{%
sv}}$ and $N_{\mathrm{cat}}$ are respectively as follows:
\begin{eqnarray}
N_{\mathrm{sv}} &=&\cosh (2g)(\left\vert \beta \right\vert ^{2}+\bar{n}%
_{s})+2\sinh ^{2}g,  \label{Nsq} \\
N_{\mathrm{cat}} &=&\cosh (2g)(\left\vert \beta \right\vert ^{2}+\bar{n}%
_{\alpha })+2\sinh ^{2}g+2\bar{n}_{q},  \label{Ncat}
\end{eqnarray}%
where the first term on the right-hand side, $\cosh (2g)(\left\vert \beta
\right\vert ^{2}+\bar{n}_{s})$ or $\cosh (2g)(\left\vert \beta \right\vert
^{2}+\bar{n}_{\alpha })$, results from the amplification process of the
input photon number and the second one $2\sinh ^{2}g$ corresponds to
amplification of the vacuum input state or the so-called spontaneous
process. The expression $N_{\mathrm{cat}}$ has an additional term, $2\bar{n}%
_{q}$, which is due to the intensity correlation of the two input fields.
When odd coherent state input [$\theta =(2k+1)\pi $ ($k=0,1,2,\cdots $)], $%
\bar{n}_{q}$ equals 0.

Fig.~2(b) shows the contour lines of $N_{\mathrm{cat}}$ as a function of $%
\theta $ and $\theta _{G}$. The maximum value of $N_{\mathrm{cat}}$ occurs
in two regions that are symmetric about $\theta =\theta _{G}=\pi $. The
parameters corresponding to the maximum value of the phase-sensitive photon
number and QFI are different, because QFI depends more on the fluctuation
and statistical properties of the input light field.

For a given $\alpha $ ($\bar{n}_{\alpha }=\bar{n}_{s}$) and $\theta
_{G}=\theta =\pi $, the sensitivity of $|0,\xi \rangle _{a}\otimes |\beta
\rangle _{b}$ and $|cat\rangle _{a}\otimes |\beta \rangle _{b}$\ input is
compared in Fig.~\ref{fig3}, which agrees with the result given by Lang and
Caves \cite{Lang1}. That is, for a coherent state input into one port, the
best state to inject into the second input port is the squeezed vacuum state.

\section{Losses}

In this section, we investigate the effects of losses on the phase
sensitivity. We use the extended model \cite{Zeng23} to calculate the QFIMs
in the presence of losses in one arm and losses in two arms.

For the enlarged system-environment state, the QFI $\mathcal{C}_{\mathrm{cat}%
}$\ is given by
\begin{equation}
\mathcal{C}=\left[
\begin{array}{cc}
C_{ss} & C_{sd} \\
C_{ds} & C_{dd}%
\end{array}%
\right] ,
\end{equation}%
where
\begin{eqnarray}
C_{ij} &=&4[\left\langle \psi \right\vert _{S}\left\langle l\right\vert
_{E}\sum_{l\mathrm{cat}}\frac{d\hat{\Pi}_{l}^{\dag }(x_{s},x_{d})}{d\phi _{i}%
}\frac{d\hat{\Pi}_{l}(x_{s},x_{d})}{d\phi _{j}}\left\vert \psi \right\rangle
_{S}\left\vert l\right\rangle _{E}  \notag \\
&&-\left\langle \psi \right\vert _{S}\left\langle l\right\vert _{E}\sum_{l}i%
\frac{d\hat{\Pi}_{l}^{\dag }(x_{s},x_{d})}{d\phi _{i}}\hat{\Pi}%
_{l}(x_{s},x_{d})\left\vert \psi \right\rangle _{S}\left\vert l\right\rangle
_{E}  \notag \\
&&\times \left\langle \psi \right\vert _{S}\left\langle l\right\vert
_{E}\sum_{l}-i\hat{\Pi}_{l}^{\dag }(x_{s},x_{d})\frac{d\hat{\Pi}%
_{l}(x_{s},x_{d})}{d\phi _{j}}\left\vert \psi \right\rangle _{S}\left\vert
l\right\rangle _{E}],  \notag \\
i,j &\in &(d,s).
\end{eqnarray}%
Because the additional freedom supplied by the environment should increase
the QFI. Therefore, $\mathcal{C}_{\mathrm{cat}}$ should be larger or equal
to $\mathcal{F}_{\mathrm{cat}}$. The relation between $\mathcal{F}_{\mathrm{%
cat}}$\ and $\mathcal{C}_{\mathrm{cat}}$ is found to be%
\begin{equation}
\mathcal{F}_{\mathrm{cat}}=\underset{\{\hat{\Pi}_{l}(x_{s},x_{d})\}}{\min }%
\mathcal{C}_{cat}\left[ |\psi \rangle ,\hat{\Pi}_{l}(x_{s},x_{d})\right] .
\end{equation}

Usually, losses in the interferometers can be modeled by adding the
fictitious beam splitters, and the lossy evolution of the field in two arms
is described by the Kraus operator $\hat{\Pi}_{l}(\phi _{s},\phi _{d})$ with
considering the phase shift. In realistic systems the photon losses are
distributed throughout the arms of the interferometer, which is described by
the parameter $\gamma $, instead of simply inserting the fictitious beam
splitters before or after the phase shift.

\subsection{Kraus operators of losses in one arm and in two arms}

Photon losses, a very usual noise, may happen at any stage of the phase
process and is modeled by the fictitious beam splitter introduced in the
interferometer arms. Firstly, we consider the photon losses in just one of
the two arms, for example arm $a$. A possible set of Kraus operators
describing the process without considering the phase shift is%
\begin{equation}
\hat{\Pi}_{l_{a}}=\sqrt{\frac{(1-\eta _{a})^{l_{a}}}{l_{a}!}}\eta _{a}^{\hat{%
n}_{a}/2}\hat{a}^{l_{a}},
\end{equation}%
where $\eta _{a}$ quantifies the photon losses of arm $a$ ($\eta _{a}=1$,
lossless case; $\eta _{a}=0$, complete absorption).

When the photon losses before or after the phase shifts, the Kraus operators
$\hat{\Pi}_{l_{a}}(\phi _{s},\phi _{d})$ including the phase factor the
general form ($\hbar =1$) is given by \cite{Escher}
\begin{eqnarray}
\hat{\Pi}_{la}(\phi _{s},\phi _{d}) &=&\sqrt{\frac{(1-\eta _{a})^{l_{a}}}{%
l_{a}!}}e^{i\phi _{d}\frac{(\hat{a}^{\dagger }\hat{a}-\hat{b}^{\dagger }\hat{%
b}-\gamma l_{a})}{2}}  \notag \\
&&\times e^{i\phi _{s}\frac{(\hat{a}^{\dagger }\hat{a}+\hat{b}^{\dagger }%
\hat{b}-\gamma l_{a})}{2}}\eta _{a}^{\hat{n}_{a}/2}\hat{a}^{^{l_{a}}},
\end{eqnarray}%
where $\gamma =0$ and $\gamma =-1$ describe the photons loss before and
after the phase shifts, respectively.

Interferometers with photon losses in both arms can be treated in a similar
way. A possible set of Kraus operators describing the process is \cite%
{Gong17}
\begin{align}
\hat{\Pi}_{l_{a},l_{b}}(\phi _{s},\phi _{d})=& \sqrt{\frac{(1-\eta
_{a})^{l_{a}}(1-\eta _{b})^{l_{b}}}{l_{a}!l_{b}!}}e^{i\phi _{d}\frac{\hat{n}%
_{a}-\hat{n}_{b}-\gamma _{a}l_{a}+\gamma _{b}l_{b}}{2}}  \notag \\
& \times e^{i\phi _{s}\frac{\hat{n}_{a}+\hat{n}_{b}-\gamma _{a}l_{a}-\gamma
_{b}l_{b}}{2}}\eta _{a}^{\frac{\hat{n}_{a}}{2}}\eta _{b}^{\frac{\hat{n}_{b}}{%
2}}\hat{a}^{l_{a}}\hat{b}^{l_{b}},
\end{align}%
where $\eta _{a}$ ($\eta _{b}$) quantifies the photon losses of arm $a$ ($b$%
). $\gamma _{a}=0$ and $\gamma _{b}=0$ ($\gamma _{a}=-1$ and $\gamma _{b}=-1$%
) describe\ the photons loss before (after) the phase shifts of arm $a$ and
arm $b$.

\subsection{Sensitivities of different input states with losses}

\begin{figure}[t]
\centerline{\includegraphics[width=0.45\textwidth,angle=0]{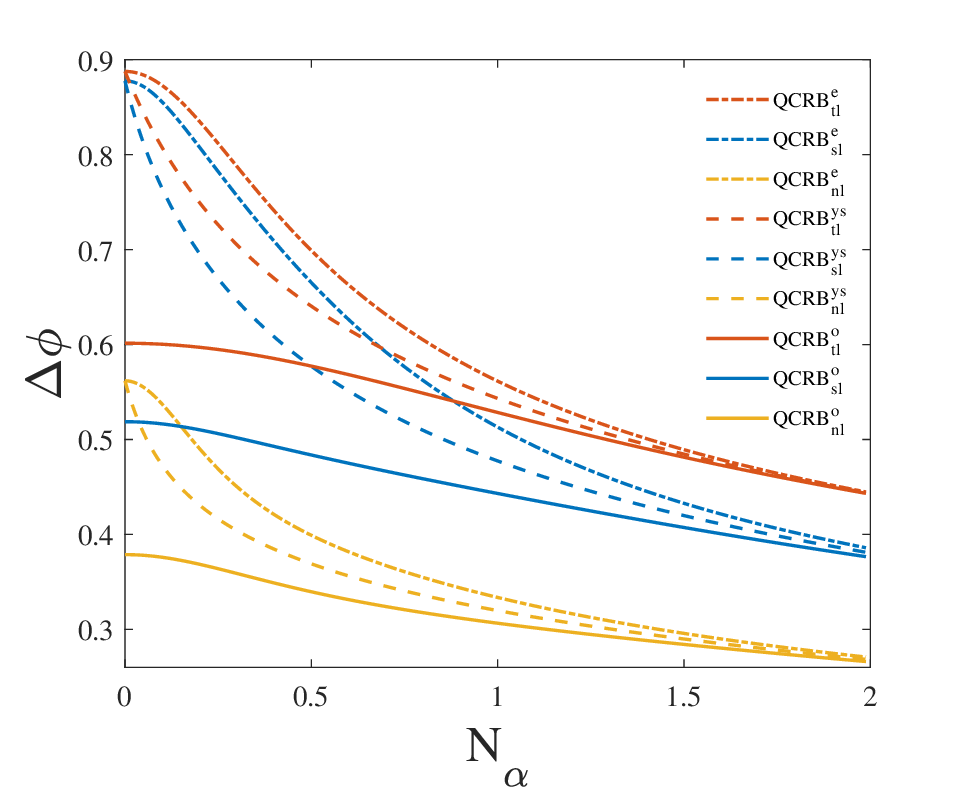}}
\caption{Phase sensitivity versus photon number $N_{\protect\alpha }$ for
different interferometer loss cases in the case of $\left\vert
cat\right\rangle _{a}\otimes \left\vert \beta\right\rangle _{b}$ input with $\left | \protect\beta \right |=0.5$, $g=1.2$, $\protect\eta%
_{a}=0.5 $ and $\protect\eta _{b}=1$ (single-arms loss
case), $\protect\eta _{a}=\protect\eta _{b}=0.5$ (two-arms loss case), and $%
\protect\theta _{G}=\pi$. QCRB$_{p}^{k}$: $k\in $[o, odd coherent states input
case; e, even coherent states input case; ys, Yurke-Stoler states input
case], $p\in $[nl, lossless case; sl, single-arm loss case; tl, two-arms
loss case].}
\label{fig4}
\end{figure}

In the case of SU(1,1) interferometers, for phase sum estimation the optimal
bound $\mathcal{C}_{\mathrm{cat}}^{\mathrm{opt}}$ for losses in one arm with
arbitrary pure state input is worked out by Zeng et al. \cite{Zeng23}, which
is given by:
\begin{equation}
\mathcal{C}_{\mathrm{cat}}^{\mathrm{opt}}=\frac{4}{\Upsilon }[\langle \hat{n}%
_{a}\rangle \frac{\eta _{a}}{1-\eta _{a}}\zeta ^{2}\langle (\Delta
n_{b})^{2}\rangle +\langle \hat{n}_{a}\rangle ^{2}(\frac{\eta _{a}}{1-\eta
_{a}})^{2}\kappa ^{2}\zeta ],\text{ \ }
\end{equation}%
where%
\begin{eqnarray}
&&\Upsilon =(\frac{\eta _{a}\langle \hat{n}_{a}\rangle }{1-\eta _{a}}%
)^{2}\kappa ^{2}(\langle (\Delta n_{a})^{2}\rangle +\langle (\Delta
n_{b})^{2}\rangle -2Cov[\hat{n}_{a},\hat{n}_{b}])  \notag \\
&&+\langle \hat{n}_{a}\rangle \frac{\eta _{a}}{1-\eta _{a}}\zeta (2\kappa
^{2}+\zeta )+\zeta ^{2}\langle (\Delta n_{b})^{2}\rangle ,
\end{eqnarray}%
and%
\begin{eqnarray}
\zeta  &=&\langle (\Delta n_{a})^{2}\rangle \langle (\Delta
n_{b})^{2}\rangle -Cov[\hat{n}_{a},\hat{n}_{b}]^{2},  \notag \\
\kappa  &=&\langle (\Delta n_{b})^{2}\rangle -Cov[\hat{n}_{a},\hat{n}_{b}].
\end{eqnarray}

The QFI for losses in two arms is given as follows:
\begin{equation}
\mathcal{C}_{\mathrm{cat}}=C_{ss}-\frac{C_{sd}C_{ds}}{C_{dd}},
\label{CQFIMP}
\end{equation}%
where
\begin{eqnarray}
C_{ss} &=&\overline{\left\langle \Delta \hat{n}_{a}^{2}\right\rangle }+%
\overline{\left\langle \Delta \hat{n}_{b}^{2}\right\rangle }+2\overline{Cov[%
\hat{n}_{a},\hat{n}_{b}]},  \notag \\
C_{dd} &=&\overline{\left\langle \Delta \hat{n}_{a}^{2}\right\rangle }+%
\overline{\left\langle \Delta \hat{n}_{b}^{2}\right\rangle }-2\overline{Cov[%
\hat{n}_{a},\hat{n}_{b}]},  \notag \\
C_{sd} &=&C_{ds}=\overline{\left\langle \Delta \hat{n}_{a}^{2}\right\rangle }%
+\overline{\left\langle \Delta \hat{n}_{b}^{2}\right\rangle },
\end{eqnarray}%
where $\overline{\left\langle \Delta \hat{n}_{a}^{2}\right\rangle }%
=[1-\Gamma _{a}(1-\eta _{a})]^{2}\left\langle \Delta \hat{n}%
_{a}^{2}\right\rangle +\Gamma _{a}^{2}(1-\eta _{a})\eta _{a}\left\langle
\hat{n}_{a}\right\rangle $, $\overline{\left\langle \Delta \hat{n}%
_{b}^{2}\right\rangle }=[1-\Gamma _{b}(1-\eta _{b})]^{2}\left\langle \Delta
\hat{n}_{b}^{2}\right\rangle +\Gamma _{b}^{2}(1-\eta _{b})\eta
_{b}\left\langle \hat{n}_{b}\right\rangle $, $\overline{Cov[\hat{n}_{a},\hat{%
n}_{b}]}=[1-\Gamma _{a}(1-\eta _{a})][1-\Gamma _{b}(1-\eta _{b})]Cov[\hat{n}%
_{a},\hat{n}_{b}]$, $\Gamma _{a}=\gamma _{a}+1$, and $\Gamma _{b}=\gamma
_{b}+1$.

In the above expression (\ref{CQFIMP}), the matrix elements $C_{ij}$ are a
function of $\Gamma _{a}$ and $\Gamma _{b}$. Then minimizing $\mathcal{C}_{%
\mathrm{cat}}$ by the parameters $\Gamma _{a}$ and $\Gamma _{b}$, the
optimal $\Gamma _{a}^{\mathrm{opt}}$ and $\Gamma _{b}^{\mathrm{opt}}$ is
obtained. Substituting $\Gamma _{a}^{\mathrm{opt}}$ and $\Gamma _{b}^{%
\mathrm{opt}}$ into $\mathcal{C}_{\mathrm{cat}}$, the minimum $\mathcal{C}_{%
\mathrm{cat}}^{\mathrm{opt}}$ is obtain, i.e., $\mathcal{F}_{\mathrm{cat}}$
in the presence of losses in two arms is achieved. However, $\mathcal{C}_{%
\mathrm{cat}}$ is more complex, and we cannot obtain an analytical solution,
which can be shown by the numerical solution according to different input
states.

\begin{figure}[t]
\centerline{\includegraphics[width=0.5\textwidth,angle=0]{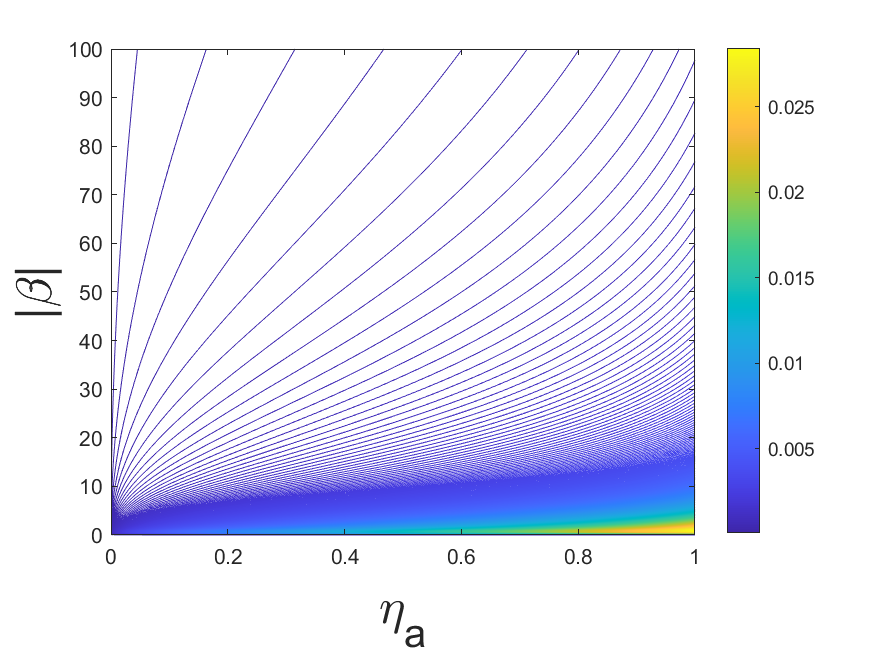}}
\caption{Phase sensitivity difference $\Delta \protect\phi _{cat-sv}$
between $\left\vert cat\right\rangle _{a}\otimes \left\vert \protect\beta %
\right\rangle _{b}$ and $\left\vert \protect\xi \right\rangle _{a}\otimes
\left\vert \protect\beta \right\rangle _{b}$ input with $\bar{n}_{\protect%
\alpha }=\bar{n}_{s}$ as a function of the photon loss factor $\protect\eta %
_{a}$ and the amplitude $\left\vert \protect\beta \right\vert $ of the
coherent state. Parameters: $\protect\alpha =2$, $g=1.2$, $\protect\theta =%
\protect\pi $ and $\protect\theta _{G}=\pi$. }
\label{fig5}
\end{figure}

When the Schr\"{o}dinger cat state and coherent state ($|cat\rangle
_{a}\otimes |\beta \rangle _{b}$) input, the average number of photons and
variances in the two arms, and the covariance are
\begin{eqnarray}
\langle \hat{n}_{a}\rangle _{\mathrm{cat}} &=&\cosh ^{2}(g)\bar{n}_{\alpha
}+\sinh ^{2}(g)(1+\left\vert \beta \right\vert ^{2})+\bar{n}_{q},  \notag \\
\langle \hat{n}_{b}\rangle _{\mathrm{cat}} &=&\cosh ^{2}(g)\left\vert \beta
\right\vert ^{2}+\sinh ^{2}(g)(1+\bar{n}_{\alpha })+\bar{n}_{q},
\end{eqnarray}%
and

\begin{eqnarray}
\langle (\Delta n_{a})^{2}\rangle _{\mathrm{cat}} &=&\sinh ^{4}(g)\left\vert
\beta \right\vert ^{2}+\sinh ^{2}(g)\bar{n}_{q}  \notag \\
&&+\frac{1}{4}\sinh ^{2}(2g)(2\bar{n}_{\alpha }\left\vert \beta \right\vert
^{2}+\bar{n}_{\alpha }+\left\vert \beta \right\vert ^{2}+1)  \notag \\
&&+\cosh ^{2}(g)(1-2\alpha ^{2}-2\bar{n}_{\alpha })\bar{n}_{q}  \notag \\
&&+\cosh ^{4}(g)(\alpha ^{4}+\bar{n}_{\alpha }-\bar{n}_{\alpha }^{2})+\bar{n}%
_{Q}-\bar{n}_{q}^{2},  \notag \\
\langle (\Delta n_{b})^{2}\rangle _{\mathrm{cat}} &=&\cosh ^{4}(g)\left\vert
\beta \right\vert ^{2}+\cosh ^{2}(g)\bar{n}_{q}  \notag \\
&&+\frac{1}{4}\sinh ^{2}(2g)(2\bar{n}_{\alpha }\left\vert \beta \right\vert
^{2}+\bar{n}_{\alpha }+\left\vert \beta \right\vert ^{2}+1)  \notag \\
&&+\sinh ^{2}(g)(1-2\alpha ^{2}-2\bar{n}_{\alpha })\bar{n}_{q}  \notag \\
&&+\sinh ^{4}(g)(\alpha ^{4}+\bar{n}_{\alpha }-\bar{n}_{\alpha }^{2})+\bar{n}%
_{Q}-\bar{n}_{q}^{2},  \notag \\
cov(\hat{n}_{a},\hat{n}_{b})_{\mathrm{cat}} &=&\cosh (2g)(1-\bar{n}_{\alpha
}-\alpha ^{2})\bar{n}_{q}+\bar{n}_{Q}-\bar{n}_{q}^{2}  \notag \\
&&+\frac{1}{4}\sinh ^{2}(2g)[\alpha ^{4}+2\left\vert \beta \right\vert ^{2}+1
\notag \\
&&+\bar{n}_{\alpha }(2+2\left\vert \beta \right\vert ^{2}-\bar{n}_{\alpha
})].
\end{eqnarray}%
Similarly, for vacuum squeezed state and coherent state ($|\xi \rangle
_{a}\otimes |\beta \rangle _{b}$) input we have%
\begin{eqnarray}
\langle \hat{n}_{a}\rangle _{\mathrm{sv}} &=&\cosh ^{2}(g)\sinh ^{2}r+\sinh
^{2}(g)(\left\vert \beta \right\vert ^{2}+1),  \notag \\
\langle \hat{n}_{b}\rangle _{\mathrm{sv}} &=&\cosh ^{2}(g)\left\vert \beta
\right\vert ^{2}+\sinh ^{2}(g)\cosh ^{2}r,
\end{eqnarray}%
\begin{eqnarray}
\langle (\Delta ^{2}n_{a})\rangle _{\mathrm{sv}} &=&\sinh ^{4}(g)\left\vert
\beta \right\vert ^{2}+2\cosh ^{4}(g)\sinh ^{2}r\cosh ^{2}r  \notag \\
&&+\frac{1}{4}\sinh ^{2}(2g)(\left\vert \beta \right\vert ^{2}e^{2r}+\cosh
^{2}r),  \notag \\
\langle (\Delta n_{b})^{2}\rangle _{\mathrm{sv}} &=&\cosh ^{4}(g)\left\vert
\beta \right\vert ^{2}+2\sinh ^{4}(g)\sinh ^{2}r\cosh ^{2}r  \notag \\
&&+\frac{1}{4}\sinh ^{2}(2g)(\left\vert \beta \right\vert ^{2}e^{2r}+\cosh
^{2}r),  \notag \\
cov(\hat{n}_{a},\hat{n}_{b})_{\mathrm{sv}} &=&\frac{1}{4}\sinh
^{2}(2g)(\left\vert \beta \right\vert ^{2}+\left\vert \beta \right\vert
^{2}e^{2r}  \notag \\
&&+2\sinh ^{2}r\cosh ^{2}r+\cosh ^{2}r).
\end{eqnarray}

The phase sensitivity versus photon number $N_{\alpha }$ with $|cat\rangle
_{a}\otimes |\beta \rangle _{b}$ input is shown in Fig.~\ref{fig4}. There
are three cases: two arm loss case (light green lines), single arm loss case
(blue lines) and lossless case (light yellow lines), and each case there are
also three different cat states input: even coherent states (dashed-dotted
lines), Yurke-Stoler states (dashed lines) and odd coherent states (solid
lines). It is shown that the sensitivity of odd coherent state input is
optimal with or without loss. The sensitivity decreases as the losses
increase from one arm to two arms. As the coherent states of Schr\"{o}dinger
cat states increase, the quantum enhancement of phase sensitivity decreases
and goes to $0$, where the Mandel $Q$-parameter $Q_{a_{in}}$ goes to $0$,
and the effect of Schr\"{o}dinger cat states is equivalent to coherent state.

Next, we numerically compare the loss resistance of the cat state and the
squeezed vacuum state. Considering that there is loss in the arm of cat
state and squeezed state, such as $\eta _{a}=50\%$, the phase sensitivity of
them can still exceed SQL as shown in Fig.~\ref{fig3}, where $\Delta \phi _{%
\mathrm{SQL}}=1/\sqrt{\langle \hat{N}\rangle }$ with $\langle \hat{N}\rangle
$ the total number of photons. In the presence of loss, the sensitivity of
squeezed state input is still better than cat state input, but the advantage
of squeezed state input is smaller than lossless case. The phase sensitivity
difference $\Delta \phi _{\mathrm{cat-sv}}$ as a function of $|\beta |$\ and
$\eta _{a}$\ is shown in Fig.~\ref{fig5}, where this difference between them
becomes smaller and smaller as the loss (1-$\eta _{a}$) and amplitude of the
input coherent state $|\beta |$ increase for given $\bar{n}_{\alpha }=$ $%
\bar{n}_{s}$. The numerical result $\Delta \phi _{\mathrm{cat-sv}}$ shows that the Schr\"{o}dinger cat states are more resistant to loss than
squeezed vacuum states.

\section{conclusion}

In conclusion, we theoretically study the quantum enhancement of the
parameter estimation by changing the quantum statistics of the Schr\"{o}%
dinger cat state with it input the SU(1,1) interferometer. The phase
sensitivity is dependent on the relative phase $\theta $ between two
coherent states of Schr\"{o}dinger cat states. The optimal sensitivity is
achieved when the odd coherent states input with sub-Poissonian statistics.
Compared with the squeezed vacuum state, the phase sensitivity of the odd
coherent state is inferior, but is more resistant to loss. These results
should be helpful in the practical application, such as quantum precision
measurement, quantum sensing, and so on.

\section{Acknowledgments}

This work is supported by the Innovation Program for Quantum Science and
Technology 2021ZD0303200; the National Natural Science Foundation of China
Grants No.~11974111, No.~12274132, No.~12234014, No.~11654005, No.~11974116,
No.~11874152, and No.~91536114; Shanghai Municipal Science and Technology
Major Project under Grant No.~2019SHZDZX01; Innovation Program of Shanghai
Municipal Education Commission No.~202101070008E00099; the National Key
Research and Development Program of China under Grant No.~2016YFA0302001;
Chinese National Youth Talent Support Program; and Fundamental Research
Funds for the Central Universities. W. Z. acknowledges additional support
from the Shanghai Talent Program.

\end{document}